\documentclass{aa}
\usepackage{amsmath}
\usepackage{graphicx}
\usepackage{amssymb}

\newcommand{\dt}[1]{\ensuremath{\frac{\partial #1}{\partial t}}}

\newcommand{\dtt}[1]{\ensuremath{\frac{\partial^2 #1}{\partial t^2}}}
\newcommand{\ds}[1]{\ensuremath{\frac{\partial #1}{\partial s}}}
\newcommand{\dss}[1]{\ensuremath{\frac{\partial^2 #1}{\partial s^2}}}

\newcommand{\paral}{\ensuremath{/\!\!/}} 
\newcommand{\para}[1]{\ensuremath{{#1}_{{{\!\text{\tiny\paral}}}}}}
\newcommand{\0}[1]{\ensuremath{{#1}_{\!o}}} 
\newcommand{\per}[1]{\ensuremath{\delta\!#1}} 
\newcommand{\db}{\ensuremath{\per{\vec{B}}}} 
\newcommand{\dT}{\ensuremath{\per{\vec{T}}}} 
\newcommand{\dbn}{\ensuremath{\per{\vec{b}}}} 
\newcommand{\dbpn}{\ensuremath{\per{\vec{b}_\perp}}} 
\newcommand{\dv}{\ensuremath{\per{\vec{v}}}} 
\newcommand{\bo}{\ensuremath{\0{\vec{B}}}} 
\newcommand{\ep}{\ensuremath{\para{\vec{e}}}} 
\newcommand{\el}{\ensuremath{\vec{e}_l}} 
\newcommand{\ea}{\ensuremath{\vec{e}_{\!A}}} 
\newcommand{\kpa}{\ensuremath{\para{k}}} 
\newcommand{\vkpe}{\ensuremath{\vec{k}_\perp}} 
\newcommand{\kpe}{\ensuremath{k_\perp}} 
\newcommand{\vxi}{\ensuremath{\vec{\xi}}} 
\newcommand{\xip}{\ensuremath{\para{\xi}}} 
\newcommand{\xil}{\ensuremath{\xi_l}} 
\newcommand{\xia}{\ensuremath{\xi_A}} 
\newcommand{\kc}{\ensuremath{\vec{\mathcal{K}}_{c}}} 
\newcommand{\kb}{\ensuremath{\vec{\mathcal{K}}_{b}}} 
\newcommand{\kbp}{\ensuremath{\vec{\mathcal{K}}_{b_{\!\perp}}}} 
\newcommand{\kr}{\ensuremath{\vec{\mathcal{K}}_{\rho}}} 
\newcommand{\gp}{\ensuremath{\mathbf{\Gamma}^{{}^{\!+}}\!}} 

\newcommand{\gm}{\ensuremath{\mathbf{\Gamma}^{{}^{\!-}}\!}} 

\newcommand{\gmaa}{\ensuremath{\mathbf{\Gamma}^{{}^{\!-}}\!}_{{}_{\!\!\!\!\mathbf{AA}}}}
\newcommand{\gmal}{\ensuremath{\mathbf{\Gamma}^{{}^{\!-}}\!}_{{}_{\!\!\!\!\mathbf{Al}}}}
\newcommand{\gmll}{\ensuremath{\mathbf{\Gamma}^{{}^{\!-}}\!}_{{}_{\!\!\!\!\mathbf{ll}}}}
\newcommand{\gmla}{\ensuremath{\mathbf{\Gamma}^{{}^{\!-}}\!}_{{}_{\!\!\!\!\mathbf{lA}}}}
\begin{document}
\thesaurus{02 
           (02.13.3; 
            09.10.1; 
            11.10.1) 
           }
\title{Pressure- and magnetic shear- driven instabilities in rotating MHD jets}
\author{Evy Kersal\'e\inst{1}
         \and Pierre-Yves Longaretti\inst{1}
         \and Guy Pelletier\inst{1,2}}
\offprints{P.-Y. Longaretti}
\institute{Laboratoire d'Astrophysique,
           Observatoire de Grenoble,
           BP 53X, Grenoble Cedex 9, F38041, FRANCE\\
           e-mail: [forename].[name]@obs.ujf-grenoble.fr
     \and
           Institut Universitaire de France\\
           }
\date{Received XXX ; Accepted XXX}
\maketitle
\begin{abstract}
 We derive new stability criteria for purely MHD 
instabilities in rotating jets, in the framework 
of the ballooning ordering expansion. Quite unexpectedly, they involve 
a term which is linear in the magnetic shear. This implies 
that cylindrical configurations can be destabilized by a negative magnetic 
shear as well as by a favorable equilibrium pressure gradient, in distinction
with the predictions of Suydam's stability criterion, which suggests on the
contrary that the shear is always stabilizing. 

  We have used these criteria to establish sufficient conditions for 
instability. In particular, the magnetic shear can always destabilize
jets with vanishing current density on the axis, a feature which is 
generically found in jets which are launched from an accretion disk.
We also show that standard nonrotating jet models (where the toroidal 
field dominates the poloidal one), which are known to be unstable,
are not stabilized by rotation, unless the plasma $\beta$ 
parameter and the strength of the rotation forces are both close to the 
limit allowed by the condition of radial equilibrium. 

The new magnetic shear-driven instability found in this paper, as well as
the more conventional pressure-driven instability, might
provide us with a potential energy source for the particle acceleration 
mechanisms underlying the high energy emission which takes place in the  
interior of AGN jets.

\end{abstract}

\section{Introduction}

  Jets are commonly observed in connection with Active Galactic Nuclei (AGNs)
and Young Stellar Objects (YSOs). These jets are cylindrically collimated over 
remarkably long distances in comparison with their radial extents; the
confinement provided by the magnetic tension of the toroidal field in current 
carrying jets has long been recognized as an efficient way to achieve this
type of collimation (as shown, e.g., by Chan and Henriksen 1981, Blandford and 
Payne 1982, after an initial suggestion by
Benford 1978). However, it is well-known from the field of thermonuclear
fusion that toroidally confined plasma configurations are generically 
unstable, and instabilities in this context are able to destroy these
configurations on very short time-scales (for a general background on MHD
instabilities, in particular in the context of thermonuclear fusion, see
Bateman 1978, and  Freidberg 1987). Furthermore, jets are subject to other
instabilities, most notably the Kelvin-Helmholtz one, which pose similar
threats to their survival.
On the other hand, there is ample evidence of instability in the observed 
jets, both directly (lateral displacements of the jet beam, bright 
knots...) and indirectly (models of jet synchrotron emission rely on particle
acceleration through e.g. shock waves or turbulence), but the observed jet survival implies that (for reasons which are still unclear), instabilities in real
jets lead to internal reorganisation and turbulence rather than to disruption.  

  Most of the literature on jet stability has focused on the Kelvin-Helmholtz
instability (see, e.g., Birkinshaw 1991 and references therein), both
for hydrodynamical and MHD jets, and for a variety of jet equilibria. Two
types of modes are produced by this instability: ordinary surface modes and
reflected body modes. Surface waves are effectively confined to the jet 
interface with the external medium, while body modes appear and become 
dominant only for beam velocities in excess of the fast magnetosonic velocity
(up to a factor of order unity). Modes growth rates decrease when the  
Mach number increases. A longitudinal magnetic field has a stabilizing 
influence due to magnetic tension; in particular, sub-Alfv\'enic (up to a
factor of order unity) flows are completely stabilized. Radiative effects 
can either enhance or reduce the growth rate of the instability depending 
on the steepness of the temperature dependence of the cooling function; in 
the presence of radiative cooling, the surface waves are apparently the most 
dangerous for jet survival (Hardee and Stone 1997; Stone {\it el al.} 1997).

  Besides the Kelvin-Helmholtz instability, driven by velocity gradients, 
jets can be unstable with respect to purely MHD processes. In the context 
of ideal MHD, these instabilities are usually divided into pressure-driven 
and current-driven instabilities (see e.g. Freidberg 1987 for details). 
Pressure-driven instabilities are related to gradients of the 
equilibrium pressure and to magnetic field line curvature. The excited
modes include the so-called saussage mode, and are usually divided into 
interchange modes and ballooning modes; they
share some common features with the Rayleigh-Taylor and Parker 
instabilities. Current-driven modes originate in the current parallel 
to the magnetic field, 
and include the so-called kink instabilities; the most dangerous 
is usually the $m=1$ kink mode ($m$ being the azimuthal wave-number). 

  Much less attention has been devoted to the analysis of these MHD
instabilities, most probably because in superfast jets (whose beam velocity 
exceeds the fast magnetosonic one), the kinetic energy exceeds the magnetic 
energy, and the Kelvin-Helmholtz instability is expected to dominate. Indeed, 
this expectation is borne out both in superfast and transmagnetosonic jets
(superalfvenic, but subfast beam velocities) for current-driven instabilities, 
whose growth rates are always substantially
smaller than for the Kelvin-Helmholtz instability, at least for force-free
jet equilibrium configurations (Appl and Camenzind 1992; Appl 1996). 
These authors also show that the jet current is stabilizing for 
supermagnetosonic flows, whereas it is destabilizing for transmagnetosonic 
ones.  
 
  This paper focuses on pressure- and magnetic shear-driven instabilities, 
in the framework of the ballooning ordering expansion. Pressure instabilities
have been most actively studied in fusion research (Kadomtsev 1966;
Coppi {\it et al.} 1979; 
Dewar and Glasser 1983; see Freidberg 1987, and references therein). In
astrophysics, they have been considered in the context of
solar (e.g., Hood, 1986) and magnetospheric physics (see Ferri\`ere 
{\it et al.} 1999 for a recent and synthetic overview 
of the situation in this field). Although these instabilities have been 
virtually ignored in the context of MHD jets (with the exception of Begelman 
1998), it is important to characterize their conditions onset there; indeed,
in spite of their local nature, they are able to produce large-scale 
disruptions over a dynamical time-scale in fusion devices. Furthermore, 
our analysis is also motivated by the following
considerations. Small-scale Kelvin-Helmholtz and current-driven instabilities 
are stabilized by magnetic tension. Furthermore, the Kelvin-Helmholtz 
instability in the transmagnetosonic regime is confined to the jet boundary. 
These features don't make them favorable energy sources for the high energy 
emission which takes place in jet interiors. By contrast, the pressure- and 
magnetic shear-driven instabilities considered here usually have 
a small-scale component transverse to the magnetic field lines, and are 
expected to occur in the type of magnetic configurations which are typically 
considered for jet inner regions. 

  In this paper, we clarify somewhat
the relation between the various classical stability criteria found in the
literature, and we assess the role of the magnetic shear and of
the rotation of the jet on the onset of jet instability, 
in particular in jet interiors. In the process, we point out the possible
and unexpected destabilizing role of the magnetic shear. This paper 
is organized as follows. In Section 2, we recall the MHD
equations and introduce our notations; we also briefly recall the origin 
of the ballooning ordering expansion, and present the perturbation equations 
in this approximation (their derivation is performed in the Appendix); finally,
we introduce the type of modes we analyse in this paper (which are interchange
and not ballooning modes) and specialize 
the ballooning equations to cylindrical 
systems. In Section 3, we focus on the mode dispersion relation and discuss 
its properties for non-rotating and rotating jets. Section 4 discusses 
our results and concludes this paper.

\section{MHD equations and ballooning ordering}
  
  For simplicity, we consider cylindrically symmetric jets, and we neglect
the shear of both the vertical and angular velocities. We
do not consider the question of the radial jet boundary. Quantities pertaining
to the jet equilibrium configuration are labelled with a ``0" subscript.

\subsection{MHD equations for jet equilibrium and perturbation}

In a frame which is rotating and moving with the jet, the MHD momentum 
equation reads

\begin{equation}
  \dt{\vec{v}} = - \vec{v} \!\cdot\! \nabla \vec{v} - \frac{\nabla
    P_\ast}{\rho} + \frac{\vec{B} \!\cdot\! \nabla
    \vec{B}}{\rho\,\mu} + \Omega_o^2 r \vec{e}_o - 
    2 \vec{\Omega}_o \times \vec{v},
  \label{equ:mvt}
\end{equation}

\noindent where $\vec{v}$ is the fluid velocity in the moving frame, $P_\ast$ 
and $\rho$ the fluid total (gaseous $P$ and magnetic $P_m$) pressure and 
density, $\vec{B}$ the magnetic field, $\mu$ the vacuum permeability, $r$
the cylindrical radius and $\vec{e_r}$ the radial unit vector. This equation 
is closed as usual with the mass continuity equation, the induction equation 
in the flux freezing approximation, and an adiabatic equation of state (we 
assume that $P\rho^{-\gamma}$ is constant throughout the plasma for simplicity).
  
  The cylindrical equilibria considered in this paper are best characterized
by introducing a number of vectors, $\kc\equiv \ep \!\cdot\!  \nabla\ep$, 
$\kb\equiv \nabla B_o/B_o$, and $\kr\equiv\nabla \rho_o/\rho_o$, where 
$\ep \equiv \vec{B}_o/B_o$ is the unit vector in the
direction of the unperturbed magnetic field $\vec{B}_o$; $\kc$ is the
curvature vector of the magnetic field lines, and $\kb$ characterizes the 
inverse of the spatial scale of variation of the magnetic field, while $\kr$
characterizes the inverse scale of variation of the fluid density. We also
introduce  the plasma $\beta$ parameter: $\beta\equiv (C_S/V_A)^2$ where 
$C_S$ and $V_A$ are the sound and Alfv\'en speed. 
This parameter measures the relative importance of the gas and magnetic 
pressures (our definition differs from the standard one by a factor 
$\gamma/2$).  With these definititions, the jet force equilibrium relation 
reads

\begin{equation}
  \rho_o V_A^2\beta\kr=\0{\rho} V_A^2\left( \kc -\kbp
  \right) + \rho_o\Omega_o^2 r \vec{e_r}.
  \label{equ:kckb}
\end{equation}

\noindent This equation implies in particular that the component of $\kr$ 
parallel to the magnetic field vanishes.

  Linearizing the momentum equation Eq.~(\ref{equ:mvt}), we obtain

\begin{multline}
 \label{equ:mvtlin} 
 \frac{\partial \dv}{\partial t} 
  =-\frac{\nabla \per{P_\ast}}{\0{\rho}} + \frac{(\bo \!\cdot\!
    \nabla) \db}{\0{\rho}\,\mu} + \frac{(\db \!\cdot\! \nabla)
    \bo}{\0{\rho}\,\mu}\\
    + \delta\rho\Omega_o^2 r\vec{e_r}-2 \vec{\Omega}_o \times \dv.
\end{multline}

\noindent where $\dv$, $\per{P_\ast}$, $\delta\rho$ and $\db$ are the 
perturbed velocity, total pressure, density,  and magnetic field, 
respectively. In terms of the perturbed gas pressure, one has 

\begin{equation}
  \per{P_\ast} = \per{P} + \frac{\bo \!\cdot\! \db}{\mu}.
  \label{equ:press}
\end{equation}

\noindent For our purposes, the linearized momentum equation is most useful
in lagragian form. Introducing the lagragian displacement $\vxi$, such that
$\vec{v}=\partial\vxi/\partial t$, the continuity and induction equations
can be integrated to yield

\begin{align}
  \label{equ:xirho} \per{\rho} &
  =-\nabla \!\cdot\!(\0{\rho} \,\vxi),\\[0.3\baselineskip]
  \db &
  =\nabla \!\times\!(\vxi \!\times\! \bo)\nonumber\\
  \label{equ:xib} & = -\bo \nabla \!\cdot\! \vxi
  -(\vxi \!\cdot\! \nabla)\bo
  +(\bo \!\cdot\! \nabla)\vxi,
\end{align}

\noindent while the expression of the total pressure perturbation follows
from our adiabatic equation of state:

\begin{equation}
  \label{equ:xiP} \per{P_\ast} 
  =\underbrace{-\vxi \!\cdot\! \nabla \0{P} -\gamma\, \0{P} \,\nabla
    \!\cdot\! \vxi}_{\displaystyle \per{P}} + \underbrace{ \frac{\bo
      \!\cdot\! \db}{\mu}}_{\displaystyle \per{P_{\text{m}}}},
\end{equation}
\noindent where $\delta P$ and $\delta P_{\rm m}$ represent the gas and 
magnetic pressure perturbations, respectively.

\subsection{Ballooning formalism:}

  The origin of MHD instabilities lies in inhomogeneous or non-static MHD
equilibria. In this paper, we focus on instabilities related to equilibrium
pressure-gradients, field line curvature and magnetic shear; they are amenable 
to an analytic description in the framework of the ballooning ordering
(Newcomb, 1961). As this formalism is not well-known in the astrophysical
community, we briefly describe it below, and generalize it to rotating systems
in a straightforward manner.

The rationale of the ballooning expansion scheme follows from properties of
the linearized magnetic tension and of MHD wave propagation in homogeneous 
media. Let us consider plane waves, where $\xi\propto
\exp(i \omega t - i\vec{k}\cdot\vec{r})$, and assume that the direction of
propagation is nearly perpendicular to the magnetic field, i.e. $\kpa\ll\kpe$
where $\kpa$ and $\kpe$ are the components of $\vec{k}$ parallel and 
perpendicular to the magnetic field, respectively. In this limit, the
frequencies of the 
slow [$\omega_S^2\simeq C_S^2 V_A^2/(C_S^2+V_A^2)\kpa^2$] and 
Alfv\'en [$\omega_A^2=V_A^2\kpa^2$]  modes are very small compared 
to the fast one  [$\omega_F^2\simeq(C_S^2+V_A^2)\kpe^2$] (see Appendix). 

On the other hand, inhomogeneous
equilibria are characterized by a scale $L_o$ of inhomogeneity (e.g.,
$|\kc|\sim |\kb|\sim L_o^{-1}$). Order of magnitude considerations show that
such inhomogeneities will contribute terms of order $V_A^2L_o^{-2}$ or
$C_S^2 L_o^{-2}$ to the dispersion relation, i.e. can destabilize the slow
and Alfv\'en modes inasmuch as the mode scale in the direction of the magnetic
field, represented by $\kpa^{-1}$ in homogeneous equilibria, is not 
significantly smaller than $L_o$ (we leave aside the fast mode for the
time being).  

\begin{figure}
  \centerline{\includegraphics{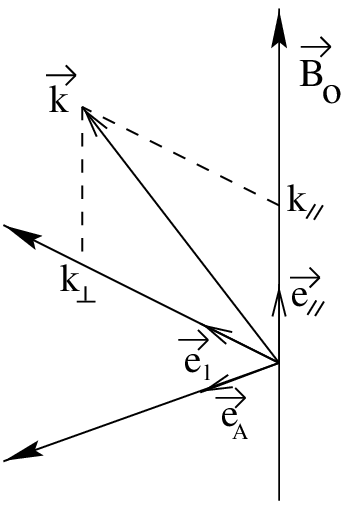}}
  \caption{Definition of the reference frame $(\ep,\el,\ea)$}
  \label{fig:triedre}
\end{figure}

With this consideration in mind, let us go back to wave propagation in 
homogeneous media, and introduce the orthogonal reference
frame ($\ep$, $\el$, $\ea$) where $\ep\equiv \vec{B}_o/B_o$ is parallel to 
the unperturbed magnetic field, $\el\equiv\vkpe/\kpe$ is parallel to 
perpendicular component of $\vec{k}$, and $\ea\equiv\ep\times\el$ (see 
Fig.~\ref{fig:triedre}); the subscripts $l$ and $A$ stand for longitudinal and
Alfv\'enic, respectively ($\ep$, $\el$, and $\ea$ are the directions of the 
displacement of purely fast, slow and Alfv\'enic modes respectively in the
limit of nearly transverse propagation adopted here). 
Denoting $(\xip,\xil,\xia)$ the components of
the lagragian displacement $\vxi$ in this reference frame, and introducing
the small parameter $\epsilon\equiv \kpa/\kpe$, the momentum equation in
the direction of $\el$ implies that $\xil\sim\epsilon\xip$ for
the slow mode; this implies in turn that $\per{P_\ast}=0$ to
leading order in $\epsilon$ (see Appendix). This reflects the fact that in 
quasi-perpendicular propagation, slow magnetosonic perturbations behave 
quasistatically and purely compressively in the fast direction (parallel 
to $\el$). This cancellation of the total pressure is essential
from a technical point of view: it allows us to introduce a WKB-like 
approximation perpendicular to the magnetic field, because it can be used
to remove all spatial derivatives of the mode amplitude in the direction 
perpendicular to field lines, as argued in the Appendix; this
greatly reduces the complexity of the problem. 
Incidentally, this eliminates fast modes from the analysis; in any case, 
these modes cannot be destabilized by the process considered here in this 
WKB-like approximation, as the terms connected to inhomogeneities are always 
subdominant in their dispersion relation. 

  In this discussion, we have focused
on instability, but the formalism can account for overstability as well, 
which can arise when considering non-static equilibria. It turns out that 
this possibility plays little role for the modes we consider (see next 
section); the question of overstability for the fast magnetosonic mode is 
considered in section 4. 

Turning back to inhomogeneous media, let us consider lagrangian displacements 
such that $\vxi=\vxi_o\exp (i\omega t - i\Phi)$. The standard ballooning 
formalism (e.g. Dewar and Glasser 1983) assumes 
$\vkpe=\nabla\Phi\perp\vec{B}_o$ everywhere, 
introduces a small parameter $\epsilon\equiv (\kpe L_o)^{-1}$, so that
variations in the direction perpendicular to the unperturbed magnetic field
are treated in the WKB approximation, but not parallel ones. As suggested by
the properties of wave propagation in homogeneous media, one introduces the
following self-consistent ordering:

\begin{align}
  \label{equ:orderomeg} \omega^2 & 
      \sim O(V_A^2 L_o^{-2},\ C_S^2 L_o^{-2}),\\[0.3\baselineskip]
  \label{equ:orderxil} \xil & 
      \sim O(\epsilon\xia,\ \epsilon\xip).
\end{align}
 
  As shown in the Appendix, the longitudinal component of the momentum
equation implies $\delta P_\ast=0$  to leading order in $\epsilon$, as 
expected, while the two remaining components reduce to:

\begin{equation}
  \dtt{\xia} - \mathcal{D}_A \!\cdot\! \xia = \mathcal{C}_A \!\cdot\!
  \xip + \mathcal{F}_{\!A}(\vxi),
  \label{equ:displ}
\end{equation}
\begin{equation}
  \dtt{\xip} -\para{\mathcal{D}} \!\cdot\! \xip = \para{\mathcal{C}}
  \!\cdot\! \xia + \para{\mathcal{F}}(\vxi),
  \label{equ:disppa}
\end{equation}

\noindent where $\cal{F}$ stands for the inertial and Coriolis forces due
to the rotation of the jet. In writing down these equations, we have defined
auxiliary quantities, $\cal{C}$ and $\cal{D}$, which characterize the 
perturbation of the magnetic tension; $\cal{C}$ introduces coupling between
the Alfv\'enic and slow magnetosonic part of the perturbation, and vanishes
for homogeneous media. Defining $\partial/\partial s\equiv \ep .\nabla$,
these quantities can be expressed as follows:

\begin{multline}
  \label{equ:dalf}
  \mathcal{D}_{A} \!\cdot\! \xia = V_A^2 \Bigg\{ \dss{}\xia +
  \mathcal{K}_{b_{\paral}} \ds{\xia} \\
      +\Biggl[ \ds{}\gmaa +
  \mathcal{K}_{b_{\paral}} \gmaa + \left(\gp\gm\right)
   _{\mbox{}_{\!\mathbf{AA}}}\\ 
  + 2 \,\mathcal{K}_{c_A}\frac{\beta}{1+\beta} \Bigl(
  \mathcal{K}_{\rho_{\!A}} - \mathcal{K}_{c_{\!A}} -
  \mathcal{K}_{b_{\!A}} \Bigr) \Biggr] \xia \Bigg\},
\end{multline}
\begin{equation}
  \label{equ:calf}
  \mathcal{C}_{A} \!\cdot\! \xip = 2 \, V_{\rm SM}^2 \,\mathcal{K}_{c_A}
  \left( \ds{\xip} - \mathcal{K}_{b_{\paral}} \xip +
    \mathcal{K}_{\rho_{\paral}} \xip \right),
\end{equation}
\noindent for the Alfv\'enic equation, and
\begin{equation}
  \label{equ:dpara}
  \para{\mathcal{D}} \!\cdot\! \xip = \frac{1}{\0{\rho}}\,\ds{}
  \left[\0{\rho} \,V_{\rm SM}^2 \,\left( \ds{\xip} - \mathcal{K}_{b_{\paral}}
      \xip  + \mathcal{K}_{\rho_{\paral}} \xip
    \right)\right],
\end{equation}
\begin{multline}
  \label{equ:cpara}
  \para{\mathcal{C}} \!\cdot\! \xia = \frac{1}{\0{\rho}}\,\ds{} \left[
    \0{\rho} \,V_{\rm SM}^2 \,\Big(\mathcal{K}_{\rho_A} - \,\mathcal{K}_{c_A}
    - \mathcal{K}_{b_A} \Bigr) \xia \right]\\ 
  + V_A^2 \left[ \left( \kb - \kc \right) \!\cdot\! \gm \!\cdot\!
    \vxi_{A} + \Big( \mathcal{K}_{b_A} - \mathcal{K}_{c_A} \Big)
    \ds{\xia} \right],
\end{multline}
\noindent for the slow magnetosonic equation; $V_{\rm SM}^2\equiv
\beta V_A^2/(1+\beta)$ is the slow magnetosonic speed. The matrices 
$\gp$ and $\gm$
are defined by

\begin{align}
  \Gamma^{{}^{\!\pm}}\! \!\cdot\! \vxi_{\!\perp} & = \left( \xil
    \,\ds{\el} + \xia \,\ds{\ea} \right) \pm \left( \vxi_{\!\perp}
    \!\cdot\! \nabla \right) \ep\nonumber\\[0.3\baselineskip]
        & = \sum_{\alpha=\{l,a\}} \xi_{\alpha}
  \left[ \left( \ep \!\cdot\! \nabla \right) \vec{e}_\alpha \pm \left(
    \vec{e}_\alpha \!\cdot\!  \nabla \right) \ep \right].
  \label{equ:gpm}
\end{align}
  
  In homogeneous media, only the first term of $\cal{D}_A$ and 
$\para{\mathcal{D}}$ remain. In this
case, Eqs.~(\ref{equ:displ}) and (\ref{equ:disppa}) reduce to the dispersion
relation of the Alfv\'en wave and slow magnetosonic wave, respectively. The
coupling coefficient ${\cal{C}}_A\propto {\mathcal{K}}_{c_{\!A}}$, so that the 
coupling of the
slow magnetosonic wave to the Alfv\'en wave is directly related to field line
curvature in the absence of rotation. The matrices $\Gamma^{{}^{\!\pm}}$ 
encapsulate purely geometric
effects due to the orientations of the field lines and of the wavevector
$\vec{k}$. 

  These equations are usually written in a more compact form (see
e.g. Dewar and Glasser 1983), which can be recovered after some lengthy
algebra by inserting the equilibrium relation for non-rotating jets and 
rescaling the displacement components as to include the purely geometric 
effects just pointed out. The expanded version presented here is more 
suitable for our purposes.

\begin{figure}
  \centerline{\includegraphics[scale=0.7]{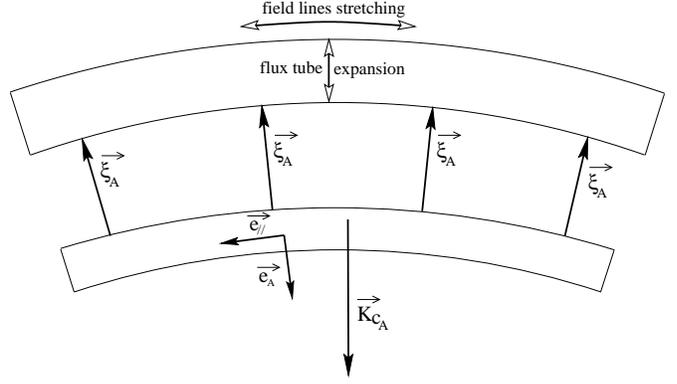}}
  \caption{Magnetic flux tube displacement}
  \label{fig:tube}
\end{figure}

  Generically, pressure-driven instabilities work on a 
combination of the alfvenic
and slow magnetosonic displacements, however, one can devise situations
where one or the other dominates. As an example, let us consider the origin 
of the destabilisation of the Alfv\'en mode in 
a static shearless two-dimensional equilibrium with no 
variation along field lines (it is also possible to devise situations where
one primarily destabilizes the magnetosonic mode, but the reasoning is less 
direct). Let us chose $\ea$ in the direction of the field 
line curvature, and consider displacements of a flux tube in the alfvenic 
direction, and constant along field lines [see fig.~(\ref{fig:tube})]. In this 
approximation, only Eq.~(\ref{equ:displ}) remains, and reduces to

\begin{equation}
    \label{equ:modealf}
    \frac{\partial^{2}\xia}{\partial t^{2}}=V_{A}^{2}\left(2\beta
    {\mathcal{K}}_{c_{\!A}}{\mathcal{K}}_{\rho_{\!A}}-4\frac{\beta}{1+\beta}
    {\mathcal{K}}_{c_{\!A}}^{2}\right)\xia.
\end{equation}

  Instability follows if the right-hand side member of this equation 
amplifies the motion, i.e. if 

\begin{equation}
    \label{crit:alf}
   {\mathcal{K}}_{\rho_{\!A}}>\frac{2}{1+\beta}
   {\mathcal{K}}_{c_{\!A}}.
\end{equation}

  This criterion reduces to the well-known criterion of instability 
with respect to $m=0$ modes (known as the ``saussage" mode, which is 
a pressure-driven mode) in cylindrical systems,  
(Kadomtsev 1966; see next subsection); it can be more directly 
established in the following way.

  Because the total pressure perturbation cancels, stability with 
respect to the applied displacement is controlled by the sign of the magnetic
tension, which, in this case, reduces to 
$\delta T_{A}=2\rho_{o}V_{A}^{2}\para{\per{b}}{\mathcal{K}}_{c_{\!A}}$ because 
the perpendicular perturbation of the magnetic field vanishes for such 
displacements. In this type of displacement, 
$\nabla.\vec{\xi}=\nabla.\vec{\xi}_{\perp}$. The considered 
displacement of the flux tube is accompanied by a compression or an 
expansion (i.e. a change in $\nabla.\vec{\xi}_{\perp}$), whose sign and 
magnitude depends on properties of the considered equilibrium, and 
which can be computed from Eq.~(\ref{equ:adbpa}), combined with 
Eqs.~(\ref{equ:kckb}) and (\ref{equ:xiP}). This procedure finally 
yields $\para{\per{b}}=\beta[{\mathcal{K}}_{\rho_{\!A}}-
2{\mathcal{K}}_{c_{\!A}}/(1+\beta)]\xia$, whose sign leads to a 
destabilizing contribution of the magnetic tension when the 
criterion (\ref{crit:alf}) is satisfied. This reasoning is typical
of an interchange instability, and in fact, the ballooning formalism
applies to both interchange and ballooning modes. 

\subsection{Reduction to cylindrical symmetry}

  The stability properties of the ballooning system of equation has
been been largely studied for cylindrical static equilibria in the context
of thermonuclear fusion research. In particular, the well-known Suydam 
criterion provides a necessary condition for stability of the ballooning 
modes (see Freidberg 1987, chapter 10 and references therein). 

  In this paper, we follow a different route by relaxing the eikonal condition 
$\vkpe=\nabla\Phi$. Instead, we Fourier transform the displacement vector
in the vertical and azimuthal directions: $\vec{\xi}=\vec{\xi}_o(\vec{r})
\exp(i\omega t - im\theta-ikz)$, i.e. we consider interchange, rather than
ballooning modes. Ballooning and interchange stability analyses in static 
media based on the form of the energy principle derived by Furth {\it et al.} 
1965 show that the destabilizing term is $\propto \kr.\kc$. For cylindrical 
configurations, this is maximized by chosing $\ea$ in the radial direction; 
this choice also leads to the simplest calculations. Therefore we assume 
$\ea=-\vec{e_r}$ from now on (the minus sign is arbitrary, and chosen so 
that important quantities appearing later are positive). This completely 
specifies the reference frame ($\ep$, $\el$, $\ea$) introduced previously. 
The existence of modes which self-consistently satisfy these constraints in
the framework of the ballooning expansion is verified in section 3.

Let us define $\vec{k}= m\vec{e}_\theta/r +k\vec{e}_z$ and 
$\kpa=\vec{k}.\vec{B}_o/B_{o}$.
The constraint of quasi-perpendicular propagation implies that $\kpa\ll
\vec{k}.\el$; this is satisfied in particular in the vicinity of magnetic 
resonances, implicitely defined by $mB_{\theta}/r+kB_z=0$ (so that 
$\kpa=0$ on resonant surfaces),
where $B_\theta$ and $B_z$ are the azimuthal and vertical components of the
unperturbed magnetic field. In general, due to magnetic shear, the constraint
of quasi-perpendicular propagation implies that our system of equations is
only valid within a region of limited radial extent for any given mode. 
However, we are mostly interested in generic 
conditions for instability rather than on mode structure; this makes 
this restriction of little
importance, since $\omega$ is constant throughout the radial extent of the 
mode, and not only in the region of validity of our system of equations, so
that conditions over $\omega$ established in this region are 
nonetheless valid for the
whole mode. Furthermore, one can define a resonant $k$ for any $m$ 
and for an arbitrary
magnetic surface, so that our analysis applies to the whole structure.
Finally, Goedbloed and Sakanaka (1974)
have proved an important theorem which implies that, at fixed $m$ and $k$, if
there is a (radially) local unstable mode, there must be a globally unstable
one with maximal growth rate for this choice of $m$ and $k$; this theorem
applies to static equilibrium configurations, i.e. in our case, to nonrotating
cylindrical columns. Its relevance for real jets is unclear, due to 
velocity shear and rotation, but it nevertheless suggests that local 
stability results (in radius) might be relevant outside their strict 
domain of validity.

  We conclude from this discussion that although the magnetic shear 
limits in theory the applicability of our analysis to a finite radial 
region for each mode, this limitation is of limited practical importance. 
Note furthermore that we have made no assumption concerning the radial
behavior of the modes we consider, besides the limitation imposed by the
nearly transverse propagation constraint, which in general should limit
both amplitude and phase variations to scales much larger than $\kpa^{-1}$ 
(however, see the discussion of section 3 for an important exception to this
rule).

  With our choice of reference frame, and the help of the equilibrium
relation Eq.~(\ref{equ:kckb}), the system of equations derived in the 
preceding section reduces to 

\begin{multline}
  \left(\omega^2-V_{\rm SM}^2\kpa^2\right)\xip=\\
   -i\left(2 V_{\rm SM}^2 \mathcal{K}_{c_{\!A}}
     \kpa+\frac{\Omega_o^2 r^2}{1+\beta}\frac{\kpa}{r}
     +2\Omega_o\omega\rho_o\frac{B_\theta}{B_o}\right)\xi_A,
  \label{equ:slow}
\end{multline}

\begin{multline}
  \left[\omega^2-(\kpa^2+k_o^2+k_r^2) V_A^2\right]\xi_A=\\
    i\left(2V_{\rm SM}^2
    \mathcal{K}_{c_{\!A}}\kpa+
    \frac{\Omega_o^2 r^2}{1+\beta}\frac{\kpa}{r}
    +2\Omega_o\omega\frac{B_\theta}{B_o}\right)\xip.
  \label{equ:alfven}
\end{multline}

\noindent In these equations, 

\begin{align}
k_r^2 &= \frac{\Omega_o^2 r}{(1+\beta) V_A^2}(\mathcal{K}_{c_{\!A}}+
      \mathcal{K}_{b_{\!A}}-\mathcal{K}_{\rho_{\!A}})
\nonumber \\[0.3\baselineskip]
&= \frac{2{\cal K}_{c_{\! A}}}{1+\beta}
\left(\frac{\Omega_o^2 r^2}{V_A^2}\right)\frac{1}{r}
  \left[1-\left(\frac{(1+\beta){\cal K}_{\rho_A}}{2{\cal K}_{c_A}}\right)
\right]\nonumber\\ & -\frac{\Omega_o^2 r^2}{(1+\beta)V_A^2}\frac{1}{r^2}
\left(\frac{\Omega_o^2r^2}{V_A^2}\right);
\label{equ:krrot}
\end{align}

\noindent $k_o$ is defined in terms of the following auxiliary quantities

\begin{equation}
\frac{h}{2\pi r}=\frac{B_z}{B_\theta},
\label{equ:h}
\end{equation}

\begin{equation}
\gp\gm=-\frac{2}{r^2}\frac{B_z^2B_\theta^2}{B_o^4}\frac{rh'}{h},
\label{equ:gpgm}
\end{equation}

\noindent and reads
\begin{align}
k_o^2 &= -\gp\gm+\frac{2\beta}{1+\beta}(\mathcal{K}_{c_{\!A}}+
       \mathcal{K}_{b_{\!A}}-\mathcal{K}_{{\rho}_{\!A}})\mathcal{K}_{c_{\!A}}.
       \nonumber\\[0.3\baselineskip]
      &= 2{\cal K}_{c_{\! A}}^2\left(\frac{B_z^2}{B_\theta^2}\right)
     \left(\frac{d \ln h}{d\ln r}\right)+\nonumber\\[0.3\baselineskip]
      & 4\frac{\beta {\cal K}_{c_{\! A}}^2}{1+\beta}
\left[1-\left(\frac{(1+\beta){\cal K}_{\rho_A}}{2{\cal K}_{c_A}}\right)
\right] -\frac{2\beta}{1+\beta}\frac{{\cal K}_{c_A}}{r}
\left(\frac{\Omega_o^2r^2}{V_A^2}\right).
\label{equ:korot}
\end{align}

\noindent In Eq.~(\ref{equ:alfven}), $V_A^2(k_o^2+k_r^2)$ plays the r\^ole of a 
generalized Brunt-V\"ais\"al\"a frequency; $k_r$ arises from the entrainment
inertial 
force term. The term $\gp\gm$ symbolizes the remaining component of the product
of $\Gamma$ matrices, and represents the effect of the magnetic shear (note
that $rh'/h=d\ln h/d\ln r$ is the magnetic shear). The
terms on the right hand side of Eqs.~(\ref{equ:slow}) and (\ref{equ:alfven})
represent the coupling between the two modes due to field line curvature,
the entrainment inertial force, and the Coriolis force, respectively. 
Note that ${\cal K}_{c_{A}}=(B_{\theta}/B_{o})^{2}/r$.

\section{Dispersion relation and instability conditions}

  The complete dispersion relation can easily be obtained from 
Eqs.~(\ref{equ:slow}) and (\ref{equ:alfven}). However, we are mostly
interested in the $\kpa\rightarrow 0$ limit, for reasons exposed below.

\subsection{Nonrotating jets}

  In order to get some grasp of the meaning of Eqs.~(\ref{equ:slow}) and
(\ref{equ:alfven}), let us first consider non-rotating jets. In this case,
the MHD perturbation equations become self-adjoint with our approximations, 
so that $\omega^2$ is real.

  Interesting general constraints can be derived from the dispersion 
relation which reads

\begin{align}
    \label{equ:disp}
    \nonumber
    \omega^{4} & -\left[\kpa^{2}(V_{A}^{2}+V_{SM}^{2})+
                   k_{0}^{2}V_{A}^{2}\right]\omega^2\\[0.3\baselineskip]
               & +\kpa^{2}(\kpa^{2}-k_{c}^{2})V_{SM}^{2}V_{A}^{2}=0,
\end{align}      

\noindent where $k_{c}^{2}\equiv \gp\gm +2\beta{\cal K}_{c_{\! A}}
{\cal K}_{\rho_{\! A}}=4\beta{\cal K}_{c_{\! A}}^{2}/(1+\beta)-k_o^2$. 
The roots can readily be extracted from this
dispersion relation, but their properties are more directly understood
in the following way.

  The product of the two roots is positive 
when $\kpa^{2}>k_{c}^{2}$. One can check that this implies that the sum is also 
positive, so that both roots are positive. Therefore 

\begin{equation}
    \label{crit:m=1}
    \kpa^{2}<k_{c}^{2}
\end{equation}

\noindent is a necessary condition of instability for the modes considered
here; it is also a sufficient condition, because the root product is negative 
then, implying that one root is negative. This relation implies in particular
$k_c^2 >0$, which, in the absence of magnetic shear, reduces to the necessary 
condition that the ballooning driving term in the energy principle of Furth 
{\it et al.} (1965) be destabilizing and is identical to (\ref{crit:alf}).
  
  Another condition is obtained by taking the limit $\kpa\rightarrow 0$ in
the dispersion relation, which yields $\omega^2=0$ and $\omega^2=V_A^2 k_o^2$
as the two roots. This implies that

\begin{equation}
    \label{crit:m=0}
    k_o^2<0
\end{equation}

\noindent is also sufficient (but not necessary) condition of instability of
the modes we consider. 

To pinpoint the meaning of the second criterion, it is instructive to examin 
the behavior of the modes when $|\kpa|\ll |k_c|,\ |k_o|$ (which can 
always be achieved by an appropriate choice of $k$ for any particular 
value of $m$). The first mode has 
$\omega^2\simeq 0$. Then, Eq.~(\ref{equ:alfven}) yields

\begin{equation}
  \label{equ:alkp=0}
  \xia\simeq\frac{-2i\beta{\cal K}_{c_A}\kpa\xip}{(1+\beta)k_o^2},
\end{equation}

\noindent which can be inserted in Eq.~(\ref{equ:slow}) to obtain

\begin{equation}
   \label{equ:omslow}
   \omega^2\simeq-V_{SM}^2\frac{k_c^2 \kpa^2}{k_o^2}.
\end{equation}

\noindent For natural reasons, we refer to the modes which exhibit this 
behavior in the $\kpa\rightarrow 0$ limit to slow (magnetosonic) modes.
From the general solution to the dispersion relation, the maximum value
of $\omega^2$ for this mode is obtained for $\kpa^2\sim k_c^2$, which
provides us with an estimate of their maximum growth rate, once inserted in 
(\ref{equ:omslow}).
  The second mode has 

\begin{equation}
    \label{equ:omalf}
    \omega^2\simeq V_A^2 k_o^2. 
\end{equation}

\noindent Then Eq.~(\ref{equ:slow}) implies

\begin{equation}
   \label{equ:slkp=0}
   \xip\simeq\frac{-2i\beta{\cal K}_{c_A}\kpa\xia}{(1+\beta)k_o^2}.
\end{equation}

\noindent We refer to modes which exhibit this behavior in the 
$\kpa\rightarrow 0$ limit as to Alfv\'en modes. The maximum growth rate for
these modes is reached for $\kpa=0$. 

  With these definitions, the slow modes are unstable and the Alfv\'en ones
stable when $k_c^2>\kpa^2$ and $k_0^2 > 0$; the situation is reversed when
$k_c^2>\kpa^2$ and $k_0^2<0$. Both modes are stabilized by the usual 
contribution to the magnetic tension when $\kpa^2>k_c^2$. 

  Note that when the magnetic shear is
important, $\xip$ for the Alfv\'en mode, and $\xia$ for the slow mode,
undergo fast radial variations. One can check that fast radial
variations of $\xip$ are not inconsistent with our procedure, 
which is not true for $\xia$. Therefore, when this occurs,
one must discard the slow mode, which is not a self-consistent solution
of our system of equations.

\subsection{Jets with vanishing current density on the axis}

  There are situations in which one expects that $B_\theta$ increases with
$r$ faster than $r$ in jet inner regions (in which case the current density
vanishes on the axis). Indeed, for jets which are launched
from an accretion disk, $u_z B_\theta/B_z =  r(\Omega-\Omega_\star)$ where
$\Omega$ and $\Omega_\star$ are the matter and field rotation respectively, 
$u_z$ is the jet velocity along the jet axis (Pelletier and Pudritz, 1992). 
In these regions, one usually expects $u_z\neq 0$, while
$\Omega, \Omega_\star\rightarrow 0$ when $r\rightarrow 0$. Furthermore,
polarisation observations indicate that the magnetic field is
predominantly aligned with the jet axis in jet cores, at least for quasar jets
(see Gabuzda, 1997, and references therein). It is therefore reasonable 
to expect that $|B_\theta|<$ or $\ll |B_z|$
over a sizeable fraction of the inner region of these jets. In these regions,
$k_0^2 < 0$ is automatically satisfied because the shear term dominates 
(as rotation is unimportant there) and
is negative [see Eq.~(\ref{equ:korot})], so that the inner regions of these
jets are generically unstable. Note that our neglect of the velocity
shear in the derivation of our equations has no impact on this conclusion,
because both rotation and velocity shear terms are expected to be 
negligible in the considered
jet regions. Obviously, $|B_z|$ must exceed $|B_\theta|$ over a sizeable
fraction of the jet interior for the instability to have noticeable effects and
growth rates.

   The destabilizing action of the magnetic shear just pointed out is the 
most important finding of this paper. It can be given a heuristic explanation
in the following way.
Regions with a destabilizing contribution of the magnetic shear are regions
in which the shear contributes to an extra-increase of the Alfv\'enic component
of the equilibrium tension when one moves outwards. Applying a purely 
radial displacement to a magnetic flux tube (i.e., considering an Alfv\'enic
mode with $\kpa=0$) produces a variation of the Alfv\'enic component of
the magnetic tension which does not include this extra piece due to the
magnetic shear; therefore, the variation in the magnetic tension produces
a restoring force on the displaced flux tube which is smaller than the 
restoring force on the equilibrium flux tube at the same location, and
instability follows.

  Note also that we must discard the magnetosonic mode, which is no longer a
valid solution of our equations due to the fast amplitude gradients of $\xia$
induced by the shear, as explained in the preceding section.

\subsection{Nonrotating jets dominated by the toroidal field}

  In magnetically confined jets, one usually assumes $|B_{\theta}|\gg |B_{z}|$
in the confining region (``Z-pinch" configurations).
This follows in particular when jets are launched from accretion
disks, because they must open considerably before the critical surfaces
are crossed, and the magnetic collimation becomes important. This
boosts the ratio $|B_{\theta}/B_{z}|$ first because the
poloidal flux is conserved within magnetic surfaces, second because
the toroidal field is considerably stretched in the process, while
this ratio is expected to be of order unity on the disk ``surface"
(Ferreira 1997; Casse and Ferreira 2000). In this case,
$\kpa\simeq m/r+kB_{z}/|B_{\theta}|$ and $k_{\perp}\simeq k$.
Furthermore $\kpa$ depends on the radial location; Eq.~(\ref{equ:slkp=0}) shows
that $\xip$ contain no large amplitude variations [as implicitly assumed
in the derivation of Eqs.~(\ref{equ:slow}) and (\ref{equ:alfven})]
provided that $|krB_{z}/B_{\theta}|\lesssim 1$ and $|m|\lesssim 1$ (note
however that fast amplitude variations of $\xia$ do not invalidate the
derivation of these equations). Then, the
condition of quasi-perpendicular propagation is satisfied for $|kr|\gg
1$ (as $m$ is a small integer). 

  When these conditions of consistency of the slow mode with our approximations
hold, one can eliminate all gradients in terms of logarithmic gradients of 
$B_\theta$ (the gradient of $B_z$ vanishes or is negligible). This allows us to
reexpress $k_o$ as

\begin{equation}
k_o^2= 4\frac{\beta {\cal K}_{c_{\! A}}^2}{1+\beta}
\left[1-\left(\frac{1+\beta}{2\beta}\right)\left(r\frac{d\ln r
B_\theta}{dr}\right)\right].
\label{equ:kB}
\end{equation}

The shear term has disappeared because it is negligible for
 $|B_\theta|\gg |B_z|$; the
first term in brackets represents the stabilizing action of the
plasma compression (it arises from the divergence of the displacement),
while the other term represents the action of pressure gradients, either
stabilizing or destabilizing. The expression for $k_{c}^{2}$ is very similar, 
and obtained by changing the sign and removing the compression term.
 
  The limit $|B_\theta|\gg |B_z|$, completed with $|kr|\gg 1$ and
$|krB_\theta/B_z| \lesssim 1$ allows us to connect with previously 
derived results on pressure-driven instabilities. In this case, one has
$\kpa={\rm sgn}(B_\theta)m/r+kB_z/|B_\theta|$, ${\cal K}_{c_A}=1/r$, 
and the criterion 
(\ref{crit:m=1}) becomes, with the help of Eq.~(\ref{equ:kB}):

\begin{equation}
   \label{crit:kadm=1}
   \frac{d\ln{|B_\theta|}}{d\ln r}>\frac{1}{2}\left(m+kr\frac{B_z}
       {B_\theta}\right)^2-1,
\end{equation}

\noindent which has been derived for $m=1$, $k=0$ modes with 
$\nabla.\vec{\xi}=0$ by Kadomtsev (1966), and for an arbitrary mode
by Begelman (1998).

  The condition (\ref{crit:m=0}) which identifies the Alfv\'en branch as the
unstable one reads in the same limit:

\begin{equation}
   \label{crit:kadm=0}
   \frac{d\ln |B_\theta|}{d\ln r}>\frac{\beta-1}{\beta+1},
\end{equation}

\noindent a condition also obtained by Kadomtsev (1966) and Begelman (1998)
for the $m=0$ ``saussage"
mode (remember that our definition of $\beta$ differs from the standard 
one by a factor $\gamma/2$). 

We point out again that (\ref{crit:kadm=1}) [or equivalently (\ref{crit:m=1})]
is a necessary and sufficient condition for having unstable modes, 
but (\ref{crit:kadm=0}) is a condition for the 
Alfv\'en branch to be unstable; these conditions do not assume any particular 
value of $m$ and do not rely on an incompressibility assumption. These
points are often overlooked in the literature. Note however that for 
smooth profiles, $k_{c}^{2}\gtrsim 1/r^{2}$, so that (\ref{crit:kadm=1}) 
limits 
$m$ to small values, independently of the limit obtained from the consistency
requirement of the slow mode. It is well-known that condition 
(\ref{crit:kadm=0}) is easy to meet in practice;
in particular, it is always satisfied for a Lorentzian (or Bennett) profile.
)
  We conclude with Begelman (1998) that the configurations usually considered
for magnetically confined jets are generically unstable in the absence of
rotation. Indeed, the condition for instability Eq.~(\ref{crit:m=1}) 
is always satisfied for $m=0$, and probably $m=1$ as well in jet inner
regions, because $|B_\theta|$ must increase from zero on the
axis to values $|B_\theta|\gg |B_z|$ in inner regions;
in general it decreases too fast in most jet models when moving 
towards the jet radial boundary for outer regions to be unstable. 
Furthermore Eq.~(\ref{crit:kadm=0}) implies 
that the Alfv\'en mode is very likely to be the unstable mode.

\subsection{Rotating jets dominated by the toroidal field}

  Rotation, in particular the Coriolis force, introduces 
an extra-coupling between
the Alfv\'enic and magnetosonic part of the displacements, which is not
vanishing in the $\kpa\rightarrow 0$ limit. Nevertheless, we stick with our
distinction between Alfv\'en and slow magnetosonic modes, although it is less
meaningful in this context.

  It is useful to characterize the jet rotation by introducing 
$\chi\equiv\Omega_o^2 r^2/
V_A^2$. The equilibrium condition Eq.~(\ref{equ:kckb}) implies that $\chi$,
like $\beta$, is of order unity at most. More specifically, one has (Pelletier 
and Pudritz 1992)

\begin{equation}
\label{equ:chi}
\chi=\frac{1}{M^2}\frac{(1-M^2 r_A^2/r^2)^2}{(1-r_A^2/r^2)^2},
\end{equation}

\noindent where $M$ is the poloidal Mach number and $r_A$ is the Alfv\'en 
radius.
In strong jets, one has $M^2 r_A^2/r^2\sim 1$ whereas $r$ is always larger
(possibly much larger) than $r_A$, which, as stated, implies that $\chi$ is 
of order unity at most. We assume $\chi\lesssim 1$, which maximizes the 
influence of rotation on the dynamics.

  The influence of the jet rotation is again analyzed by deriving the 
dispersion relation, which reads:

\begin{align}
    \omega^{4}- & \left[\kpa^{2}(V_{A}^{2}+V_{SM}^{2})+
     (k_{0}^{2}+k_r^2)V_{A}^{2}+4\Omega_o^2(B_\theta^2/B_o^2)\right]\omega^2
    \nonumber\\[0.3\baselineskip]
      & -4\Omega_o(B_\theta/B_o)\kpa V_{SM}^2
          \left(2{\cal K}_{c_A}+\frac{\chi}{\beta r}\right)\omega
          \nonumber\\[0.3\baselineskip]
      & +\kpa^{2}\bigg[\kpa^{2}-k_{c}^{2}+k_r^2 
         \nonumber\\[0.3\baselineskip]
      & -\frac{\chi}{(1+\beta)}
         \left(2{\cal K}_{c_A}+\frac{\chi}{\beta r}\right)\bigg]
           V_{SM}^{2}V_{A}^{2}=0,
    \label{equ:disprot}
\end{align}

\noindent where $k_o^2$ and $k_r^2$ are given in Eqs.~(\ref{equ:korot}) and
(\ref{equ:krrot}). 

  In the limit $\kpa=0$, this equation has two roots, the magnetosonic 
branch (whose degeneracy is broken by rotation) $\omega^2=0$, and 
the Alfv\'enic branch 

\begin{equation}
\label{equ:freqrot}
\omega^2  = V_A^2 \left(k_o^2+k_r^2\right)+4\Omega_o^2
            \left(\frac{B_\theta^2}{B_o^2}\right).
\end{equation}

\noindent The last term in this relation is due to the Coriolis
force and is always stabilizing. This is a natural consequence of the form 
of this force, $\vec{\Omega}_o\times\vec{\xi}$, which gives rise to 
coupled oscillations between the components of the displacement vector. The 
factor $B_\theta/B_o$ arises because of the inclination of the magnetic vector 
$\vec{B}_o$ with respect to the rotation axis. 

  The discussion of the stability properties in the absence of rotation implies
that stabilization by rotation will only occur when $\chi\sim 1$. To quantify
somewhat this statement, let us rewrite $k_o^2$ and $k_r^2$ as

\begin{equation}
k_o^2=\frac{1}{r^2}\frac{4\beta}{1+\beta}\left[1-\frac{1+\beta}{2\beta}
\frac{d\ln r |B_\theta|}{d\ln r}\right]+\frac{2\chi}{r^2(1+\beta)},
\end{equation}

\begin{equation}
k_r^2=\frac{2\chi}{r^2(1+\beta)}\left[1-\frac{1+\beta}{2\beta}
\frac{d\ln r |B_\theta|}{d\ln r}\right]+\frac{\chi^2}{r^2\beta(1+\beta)}.
\end{equation}

It is apparent that the contribution of the jet rotation to
$k_o^2$ is always stabilizing. Furthermore, the
pressure contributions to $k_0^2$ and $k_r^2$ are destabilizing when
condition (\ref{crit:kadm=0}) is satisfied.
The arguments of the previous subsection make this very likely to happen
in the inner region of jets. A condition on the critical gradient of $B_\theta$
can easily be extracted, but it is not very informative. Instead, limits of 
instability can be obtained 
in the following way. We assume that $d\ln |B_\theta|/d\ln r\simeq 1$; this
applies in jet inner regions when the behavior of the magnetic shear
is not controlled by rotation, as in the previous section. 
The dispersion relation (\ref{equ:freqrot}) implies then 
that $\beta\lesssim 1/4$ for $\chi\simeq 1$ or reversely $\chi\lesssim 1/2$
for $\beta\simeq 1$ constitute sufficient conditions of instability. This 
implies that it is very difficult in practice to stabilize the
Alfv\'en mode by rotation, especially that
jets possessing an important cylindrical asymptotic limit have $\chi\ll 1$
from Eq.~(\ref{equ:chi}), as such jets have $M\gg 1$.

 Note that stability occurs for {\it high} $\beta$, whereas in the 
fusion context, the reverse is true. This follows because destabilisation
is provided by the rotation entrainment force rather than pressure
forces.

 The stability properties of the slow mode can be discussed in the 
$\kpa\rightarrow 0$ limit. Assuming that the frequency of 
magnetosonic modes $\omega\propto\kpa$ for small 
enough $\kpa$, as implied by the dispersion relation, and keeping only the 
leading order term in $\kpa$ yields a reduced dispersion relation:

\begin{equation}
    A\omega^{2}+B\omega+C=0,
    \label{disp:rotmag}
\end{equation}

\noindent where

\begin{align}
    A & =
   \left[(k_{0}^{2}+k_r^2)V_{A}^{2}+
       4\Omega_o^2(B_\theta^2/B_o^2)\right],
       \nonumber\\[0.3\baselineskip]
    B & = 4\Omega_o(B_\theta/B_o)\kpa V_{SM}^2
          \left(2{\cal K}_{c_A}+\frac{\chi}
          {\beta r}\right)
          \nonumber\\[0.3\baselineskip]
    C & = \kpa^{2}\left(-k_{c}^{2}+k_r^2
       -\frac{\chi}{r(1+\beta)}
       \left(2{\cal K}_{c_A}+\frac{\chi}{\beta r}\right)\right)
           V_{SM}^{2}V_{A}^{2}.
\end{align}
 
  Because of the term $\propto\omega$ in this reduced dispersion relation,
overstability rather than instability occurs when the
discriminant $\Delta\equiv B^2 -4AC < 0$. It turns out that for $\chi\simeq 1$
and $d\ln B_\theta/d\ln r\simeq 1$, $\Delta>0$ whatever the value of $\beta$,
implying that the slow mode is stabilized by rotation in the small $\kpa$ limit.
 
  This suggests that the slow mode is more easily 
stabilized than the Alfv\'enic
one. In any case, our result on the Alfv\'en mode is sufficient to conclude
that jets with a dominant toroidal field are difficult to stabilize by 
rotation.

\section{Discussion}

  In this paper, we have focused on the stability of rotating MHD jets with 
respect to inhomogeneities connected to pressure gradients and magnetic
shear, in the framework of the ballooning ordering. The most remarkable 
finding of our analysis is the possible destabilizing 
influence of the magnetic shear pointed out in section 3. Indeed, 
the shear appears quadratically in Suydam's well-known criterion, which
applies to the stability of cylindrical columns, whereas the stability criterion
of the preceding section is linear in magnetic shear. We point out that, had
we not relaxed the eikonal condition in section 2, we would have recovered
Suydam's criterion (see Freidberg 1987, Chapter 10). This unexpected behavior
of the magnetic shear implies that the inner regions of jets launched from
accretion disks are generically unstable, as argued in section 3.

We have also recovered that
for jet models dominated by a toroidal magnetic field, jet interiors are 
likely to be subject to this type of instability, a result already pointed out
by Begelman (1998). Furthermore, we find that rotation has little influence on 
this conclusion; rotation can only stabilize jets when it is close to the
limit allowed by the radial equilibrium condition, and for values of $\beta$
close to the same limit. 

  Our results on pressure-driven instabilities agree with those found in the 
literature, most notably Begelman (1998) and Ferri\`ere {\it et al.} (1999), 
once appropriate limits are taken. Begelman's study is restricted to
$B_\theta/B_z \gg 1$, which removes the effect of the magnetic shear from the
analysis. Ferri\`ere {\it et. al.} (1999)
do not rely on the ballooning expansion, keep the full
six-dimensional dispersion relation, and discuss in detail the effects of
rotation and gravity, besides gas pressure and field curvarture. They give
a detailed and interesting synthesis of gravitational and centrifugal
instabilities
in magnetospheric studies. However, to achieve such a generality, they rely on
a particular coordinate representation of the magnetic field, which is 
well-adapted to planetary magnetospheres, but forbids also
the presence of magnetic shear.

  Jet models which are not dominated by the toroidal magnetic field have
been proposed in the literature. One example is the force-free jet model of
Appl and Camenzind, 1992, which has $B_\theta\sim B_z$, and belongs to a class
of configurations known as ``reversed field pinch" in the fusion research
context. These configurations are known to be stable with respect to 
pressure-driven instabilities for moderately low values of $\beta$ (and, more
generally, to possess good stability properties). This
apparently conflicts with our results on destabilization by the shear. 
However, we point out that destabilization occurs because $B_\theta$ varies
faster than $r$ in the jet inner region. This variation is induced by the
behavior of the jet matter and field rotation, a feature which is absent
from fusion control devices.

  In this analysis as in all others we are aware of, we have assumed that the
jet equilibrium structure is invariant along the jet axis. This need not
be the case, as examplified in various self-similar solutions (Sauty and
Tsinganos, 1994;  Contopoulos and Lovelace, 1994),
and more recenly in a complete analytic solution displaying a more complex
magnetic field structure (Bogoyalenskij, 2000). The relevance 
of the available stability
analyses to such configurations is unclear; furthermore, these configurations
are possibly unstable with respect to appropriately chosen ballooning modes.
In any case, this question should be addressed before firm conclusions can be
drawn. 

  We have not tried to derive precise growth rates, as these depend on the
radial structure of the most unstable modes, the determination of which
is outside
the scope of the approximation scheme we have adopted. However, order of
magnitude considerations show that these instabilities should develop within
a few dynamical time-scale ($\sim a V_A^{-1}$ where $a$ is the 
characteristic radial size of the considered region).

  Let us also return to the question of overstability, which was
briefly alluded to in section 2 and 3. Our dispersion relation shows 
that $\omega^2$ is real for alfvenic modes in the $\kpa=0$ limit, implying that
$\omega$ is real when $\omega^2>0$. This property remains true when 
$\omega(\kpa)$ is computed as a series in $\kpa$, to all orders in this
series if it is true to lowest order (this can be shown by induction from 
the full dispersion relation, because it
involves only real terms). Therefore the stability property of the Alfv\'en
mode is obtained in the $\kpa=0$ limit. In the same way, the dispersion 
relation of Ferri\`ere {\it et. al.} (1999) [their Eq.~(27)] implies that 
the fast magnetosonic mode is stable in the absence of magnetic shear in the
perpendicular WKB approximation. This suggests that
the fast mode is probably not destabilized in the asymptotic limit examined 
in this paper.

  We point out again that, in spite of their partly local nature, 
pressure-driven instabilities, and magnetic shear-driven ones
as well, are likely to produce 
large-scale disruptions of MHD jets. How real jets survive these instabilities
as well as Kelvin-Helmholtz and current-driven instabilities is a still
open and pressing problem, especially that 3D instabilities seem more vigorous
than there 2D counterparts (see, e.g., Bodo {\it et al.}, 1998). However,
progress on this topic has recently been accomplished. For the 
Kelvin-Helmholtz instability, recent 3D simulations have shown that
for initial magnetic Reynolds numbers smaller than $\sim 50$, the shear
layer broadens, the turbulence level decreases and the flow becomes 
quasi-laminar and stable in the end (Ryu {\it et al.}, 2000);
as jets are not expected to be drastically out of equipartition, and as
the Lorentz force provides the origin of their acceleration, this probably
ensures their stability against a possible disruption by the Kelvin-Helmholtz
instability. Also, the kink instability seems to lead to helicoidal equilibria 
with a redistribution of the current density rather than to disruption (Lery
{\it et al.}, 2000). Finally, it has recently been recognized in the
fusion literature that pressure-driven instabilities can be stabilized
by large equilibrium gradients, a somewhat counterintuitive result 
(Terry, 2000).

  One of the interesting questions raised by this work concerns the 
possible link between the type of instabilities we consider and 
the particle acceleration mechanism
which is responsible for the high energy emission seen in AGN jet interiors. 
For example, in the context of the two-flow model,
most of the high energy emission
comes from a highly relativistic pair plasma occupying the inner region of a 
slower jet which contains most of the mass and provides most of the energy.
This model
relies on energy injection to the pair plasma from the bulk of the jet through 
turbulence (Renaud and Henri, 1998); note also that the existence of a pair 
plasma in AGN jets has recently received an
interesting observational support (Wardle, {\it et. al.}, 1998; Hirotani 
{\it et. al.}, 1998). In such a picture of the high energy emission, the 
required turbulence would most likely be
provided by MHD instabilities in the nonlinear regime, as other
sources look unpromising. Indeed, a purely hydrodynamical origin in regions
of increasing angular velocity with radius, as advocated by Richard and
Zahn (1999) is not found in numerical simulations (Balbus {\it et al.}, 1996),
and happens in laboratory Couette experiments (performed in general with 
narrow gap widths between the inner and outer cylinder) only when the 
outer cylinder completely dominates the rotation (Coles, 1965), a situation
which has little connection with disks and jets; furthermore, the 
magneto-rotational instability (Balbus and Hawley, 1991) does not work in
this regime, and the velocity shear is ineffective in driving the 
Kelvin-Helmholtz instability in the 
subalfvenic portion of the jet in jet interiors.

To conclude, let us finally point out that pressure-driven modes are also
known to be avalanche-like instabilities, due to the existence of a critical 
gradient, and this feature raises
important question concerning the action of this instability on jet 
profiles, as well as their role in jet variability.

\appendix
\section{MHD perturbation equations and the ballooning approximation}
  
  Let us first recall some properties of wave propagation in homogeneous
media. In the reference frame introduced in section 2.2, the momentum equation
yields the following three component equations
\begin{equation}
  \left( \omega^2 - C_S^2\,\kpa^2 \right) \,\xip = C_S^2 \,\kpa \,\kpe \,\xil,
  \label{equ:mgs1}
\end{equation}
\begin{equation}
  \left( \omega^2 - C_S^2\,\kpe^2 - V_A^2\,k^2 \right) \,\xil = C_S^2
  \,\kpa \,\kpe \,\xip,
  \label{equ:mgs2}
\end{equation}
\begin{equation}
  \left( \omega^2 - V_A^2\,\kpa^2 \right) \,\xia = 0,
  \label{equ:alf}
\end{equation}
\noindent while the total pressure perturbation becomes
\begin{equation}
 \label{equ:dp}
    \per{P_\ast}  =-i\0{\rho} \left[ (C_S^2+V_A^2) \,\kpe\xil +
    C_S^2\,\kpa\xip \right].
\end{equation}
  Eq.~(\ref{equ:alf}) gives the dispersion relation of Alfv\'en waves, 
$\omega_A^2=V_A^2\kpa^2$, which decouple from the two magnetosonic modes 
described by the remaining two equations. In the approximation of interest 
here ($\epsilon=\kpa/\kpe\ll 1$, i.e. nearly perpendicular propagation), these 
two equations imply $\omega_S^2\simeq C_S^2 V_A^2/(C_S^2+V_A^2)\kpa^2$ and 
$\xil\sim O(\epsilon\xip)$ for the slow magnetosonic wave, while 
$\omega_F^2\simeq(C_S^2+V_A^2)\kpa^2$ and $\xip\sim O(\epsilon\xil)$ 
for the fast magnetosonic one.

  Furthermore, the $\xil$ momentum component Eq.~(\ref{equ:mgs2}) combined with
Eq.~(\ref{equ:alf}) implies that $\per{P_\ast}=0$ to leading order in 
$\epsilon$.

  Let us now turn to inhomogeneous and non-static media. The ordering assumed
in Eqs.~(\ref{equ:orderomeg}) and (\ref{equ:orderxil}) implies that 
the longitudinal component of the momentum equation [Eq.~(\ref{equ:mvt})]
reduces to

\begin{equation}
  \per{P_\ast} = \per{P} + \frac{\bo \!\cdot\! \db}{\mu}= 0.
  \label{equ:press2},
\end{equation}

\noindent to leading order in $\epsilon$. In establishing this result, the
unperturbed momentum equation [Eq.~(\ref{equ:kckb})] has been used to
show that $\Omega_o\sim \omega$ at most. The total pressure 
perturbation cancels because in the quasi-perpendicular propagation 
assumed here, slow and alfvenic perturbations behave quasistatically 
in the fast direction, and because fast perturbations are only 
compressional in the fast direction in homogeneous media, implying 
that inhomogeneous corrections are negligible in our WKB-like 
approximations. To derive the remaining two 
momentum component equations, we need to evaluate the tension perturbation
in terms of the displacement $\vxi$ [the entrainment inertial term follows 
immediately from the perturbed density Eq.~(\ref{equ:xirho})]

\begin{equation}
 \label{equ:xitens} 
 \dT=
    \frac{\bo \!\cdot\!  \nabla \db}{\mu} +
    \frac{\delta\!\vec{B} \!\cdot\!  \nabla \bo}{\mu}.
\end{equation}

  The magnetic field perturbation $\dbn = \db/\0{B}$ [cf Eq.~(\ref{equ:xib})]
along and perpendicular to the unperturbed magnetic field read 
\begin{equation}
  \para{\per{b}} = -\nabla \!\cdot\! \vxi_\perp - \left( \kbp + \kc
  \right) \!\cdot\! \vxi_\perp,
  \label{equ:adbpa}
\end{equation}
\begin{align}
  \dbpn & = \gm \!\cdot\! \vxi_{\!\perp} + \ds{\xil} \,\el + \ds{\xia}
  \,\ea + \left(\vxi \!\cdot\! \kc\right) \ep\nonumber \\[0.5\baselineskip]
        & = \begin{pmatrix}
    \displaystyle \gmll \,\xil +  \gmla \,\xia + \ds{\xil}\\[0.4cm]
    \displaystyle \gmal \,\xil + \gmaa \,\xia + \ds{\xia}
  \end{pmatrix}\ ,
  \label{equ:adbpe}
\end{align}
\noindent where the matrices $\gp$ and $\gm$ are defined in 
Eq.~(\ref{equ:gpm}), $\partial/\partial s\equiv\ep.\nabla$, and the various
$\cal K$s are defined right before Eq.~(\ref{equ:kckb}).

  In Eq.~(\ref{equ:adbpa}), the first term represents the variation 
of the field in its original direction due to the conservation of 
magnetic flux in transverse compression of the plasma; the second 
term represents the contribution of field advection, the last one the 
contribution of field line stretching.

  With these results, the perpendicular component of the tension perturbation
reads
\begin{multline}
  \per{T_{\{l,A\}}} = \0{\rho} \,V_A^2 \left( \ds{}\per{b}_{\{l,A\}} +
    \mathcal{K}_{b_{\paral}} \per{b}_{\{l,A\}}
     \right.\\[0.4\baselineskip]
     \left. + \,\vec{e}_{\{l,A\}}
    \!\cdot\! \gp \!\cdot\! \dbpn + 2 \,\per{\para{b}}
    \,\mathcal{K}_{c_{\{l,A\}}}\right),
\end{multline}

\noindent while the parallel tension perturbation is given by

\begin{equation}
  \per{\para{T}} = \0{\rho} \,V_A^2 \,\left[\dbpn \!\cdot\!
  \left( \kbp - \kc \right)+\frac{\partial \para{\per{b}}}{\partial s}
  + 2\mathcal{K}_{b_{\paral}}\para{\per{b}}\right].
  \label{equ:afl}
\end{equation}

\noindent These equations for the perturbed magnetic field and magnetic tension 
are exact (the ordering has not been used in their
derivation). Furthermore, by noting that 
\begin{equation}
\nabla \!\cdot\! \vxi_\perp = \ds{\xip} - \xip\,\mathcal{K}_{b_{\paral}} -
  \nabla \!\cdot\!  \vxi,
  \label{equ:div}
\end{equation}

\noindent it appears that derivatives of the displacement in the magnetic
force term enter only through derivative along the field line and $\nabla.\vxi$;
the same crucial feature holds for the entrainment inertial force term, while 
the derivatives of the displacement do not enter the expression of the Coriolis 
force term. With the help
of the cancellation of the total pressure perturbation Eq.~(\ref{equ:press2}), 
one can express $\nabla.\vxi$ in terms of $\vxi$ and of the derivatives of the 
components of $\vxi$ in the field line direction. This yields

\begin{align}
\nabla \!\cdot\!  \vxi= & -\frac{1}{1+\beta}\left[\left(\beta
                           \mathcal{K}_{\rho_{\! A}}+\mathcal{K}_{b_{\! A}}+
                           \mathcal{K}_{b_{\! A}}\right)\xi_A\right.
                                      \\[0.3\baselineskip]\nonumber
                        & +\beta\left(\frac{\partial}{\partial s}-
                            \mathcal{K}_{b_{\paral}}+
                            \mathcal{K}_{\rho_{\paral}}\right)\xip
                                      \\[0.3\baselineskip]\nonumber
                        & \left.-\left(1+\beta\right)
                          \left(\mathcal{K}_{b_{\! A}}+
                          \mathcal{K}_{c_{\! A}}\right)\xi_a\right]
\end{align}

\noindent This reduces the problem to 
the set of ordinary differential equations for $\xip$ and $\xi_A$ given in 
Eqs.~(\ref{equ:displ}) and (\ref{equ:disppa}).

\end{document}